\documentclass[sigconf]{acmart}

\AtBeginDocument{%
  \providecommand\BibTeX{{%
    \normalfont B\kern-0.5em{\scshape i\kern-0.25em b}\kern-0.8em\TeX}}}

\usepackage{amsthm}
\newtheorem{dfn}{Definition}

\usepackage{enumerate}
\usepackage{bm}
\usepackage{multirow}
\usepackage{multicol}
\usepackage{float}
\usepackage{caption}
\usepackage{subfigure}
\usepackage{stfloats}
\begin{document}

\title{Context-aware Heterogeneous Graph Attention Network for User Behavior Prediction in Local Consumer Service Platform}

\author{Peiyuan Zhu}
\email{xingjian.zpy@alibaba-inc.com}
\affiliation{%
  \institution{Alibaba Group}
  \city{Hangzhou}
  \country{China}
}

\author{Xiaofeng Wang}
\email{aron.wxf@alibaba-inc.com}
\affiliation{%
  \institution{Alibaba Group}
  \city{Hangzhou}
  \country{China}
}

\author{Zisen Sang}
\email{zisen.szs@alibaba-inc.com}
\affiliation{%
  \institution{Alibaba Group}
  \city{Hangzhou}
  \country{China}
}

\author{Aiquan Yuan}
\email{aiquan.yaq@alibaba-inc.com}
\affiliation{%
  \institution{Alibaba Group}
  \city{Hangzhou}
  \country{China}
}

\author{Guodong Cao}
\email{guodong.cao@alibaba-inc.com}
\affiliation{%
  \institution{Alibaba Group}
  \city{Beijing}
  \country{China}
}

\renewcommand{\shortauthors}{Zhu, et al.}

\begin{abstract}

As a new type of e-commerce platform developed in recent years, local consumer service platform provides users with software to consume service to the nearby store or to the home, such as Groupon\footnote{https://www.groupon.com} and Koubei\footnote{https://www.koubei.com}. Different from other common e-commerce platforms, the behavior of users on the local consumer service platform is closely related to their real-time local context information, such as the location or the time. Therefore, building a context-aware user behavior prediction system is able to provide both merchants and users better service in local consumer service platforms. However, most of the previous work just treats the contextual information as an ordinary feature into the prediction model to obtain the prediction list under a specific context, which ignores the fact that the interest of a user in different contexts is often significantly different. Hence, in this paper, we propose a context-aware heterogeneous graph attention network (CHGAT) to dynamically generate the representation of the user and to estimate the probability for future behavior. Specifically, we first construct the meta-path based heterogeneous graphs with the historical behaviors from multiple sources and comprehend heterogeneous vertices in the graph with a novel unified knowledge representing approach. Next, a multi-level attention mechanism is introduced for context-aware aggregation with graph vertices, which contains the vertex-level attention network and the path-level attention network. Both of them aim to capture the semantic correlation between information contained in the graph and the outside real-time contextual information in the search system. Then the model proposed in this paper aggregates specific graphs with their corresponding context features and obtains the representation of user interest under a specific context and input it into the prediction network to finally obtain the predicted probability of user behavior. After experimental verification, the proposed method in this paper not only shows superior prediction performance in large-scale offline datasets but also achieves a huge improvement in the online search click-through rate prediction experiment in the real local consumer service platform. It is worth noting that the model proposed in this paper increases the click-through rate of users by $3.85\%$ and also increases the average revenue per user by $7.99\%$, which proves that the proposed method has both great economic value and application potential in local consumer service platforms.

\end{abstract}



\maketitle







\section{Introduction}

User behavior prediction is one of the most important tasks in the current e-commerce system, where its main purpose is to predict the probability of a user taking a certain behavior toward the candidate item. It is obvious that accurate prediction improves the feeling of the experience of users and optimizes the search system's ability to precisely match materials. In general, the algorithm for modeling and predicting user behavior has a wide range of applications and has attracted lots of attention both from industry and academia.
However, when users search for products in the local consumer service platform, the local context information they are in largely affects and restricts potential interest. For example, a user may want to browse coffee and breakfast on the way to work in the morning, buy a work meal near the company at noon, and purchase fruit on the way back home. It can be seen that the change in the context would lead to a huge variation in the potential interest of users. Therefore, creating a model to accurately understand the potential interests of users under different contexts has become a key urgent in predicting user click behavior in the local consumer service platform. It can be concluded that the main challenges of modeling context-aware interest representation come from the following two aspects:
\begin{itemize}
\item How to uniformly represent the heterogeneous behavior of users and the contextual information contained in it?
\item How to dynamically generate the context-aware interest representation of users?
\end{itemize}
Next, we try to employ existing methods to solve the above challenges:
\paragraph{The User Representation Methods}
Many works have adopted user behavior sequences for representing user behaviors and modeling user’s potential interest for future click prediction, such as DIN\cite{zhou2018deep} or DIEN\cite{zhou2019deep}. However, this type of approach only makes use of information in the time dimension for interest generation, and it is difficult to distinguish the correlation between historical behavior and the current candidate items under different contexts. Moreover, some works\cite{10.1145/3292500.3330836} input artificially designed contextual features into the model to build context-sensitive predictions, but such approaches only model the cross-relationship between user interests and the context. In general, it is difficult for the existing user representation methods to naturally integrate the historical heterogeneous behaviors of users and to perform accurate user interest modeling in a frequently changing context.


\paragraph{The Graph Methods}
 From the perspective of heterogeneous behavior recording and representation, the commonly used method in the industry is to first establish a heterogeneous interest network of users and then decompose the vertices of the heterogeneous graph with tag matching\cite{10.1145/3292500.3330673}. However, such methods cannot describe contextual information and incorporate it into the process of graph aggregation. Moreover, the graph attention model is commonly used in the industry to capture vertex preferences in the graph. And most of the existing methods choose GAT or HAN as benchmarks(\cite{velivckovic2017graph},\cite{han2019}), then change the aggregation method in their respective application scenarios. But the original attention mechanism in GAT is performed on a complete graph that has been constructed and is utilized to compare its correlation with all neighbors on the graph. Leaving such attention methods poor performance to integrate real-time contextual information for graph aggregation. 

\begin{figure*}[htbp]
\centering
\subfigure[Graph Construction]{
\centering
\includegraphics[scale=0.35]{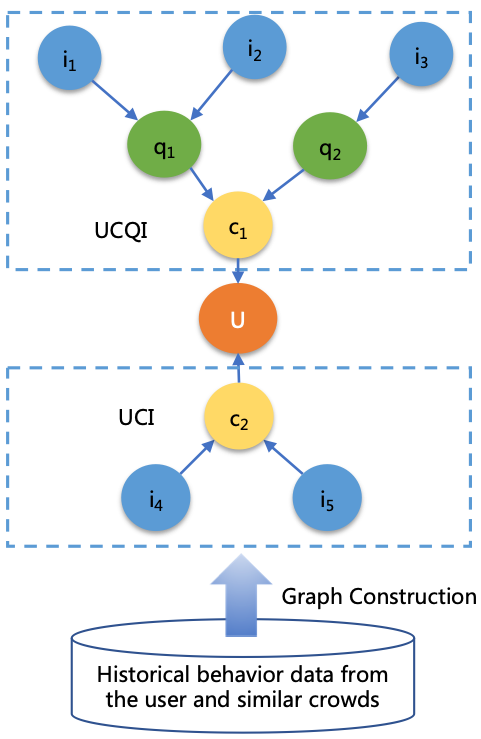}
}\hspace{0.5mm}
\subfigure[CHGAT]{
\centering
\includegraphics[scale=0.35]{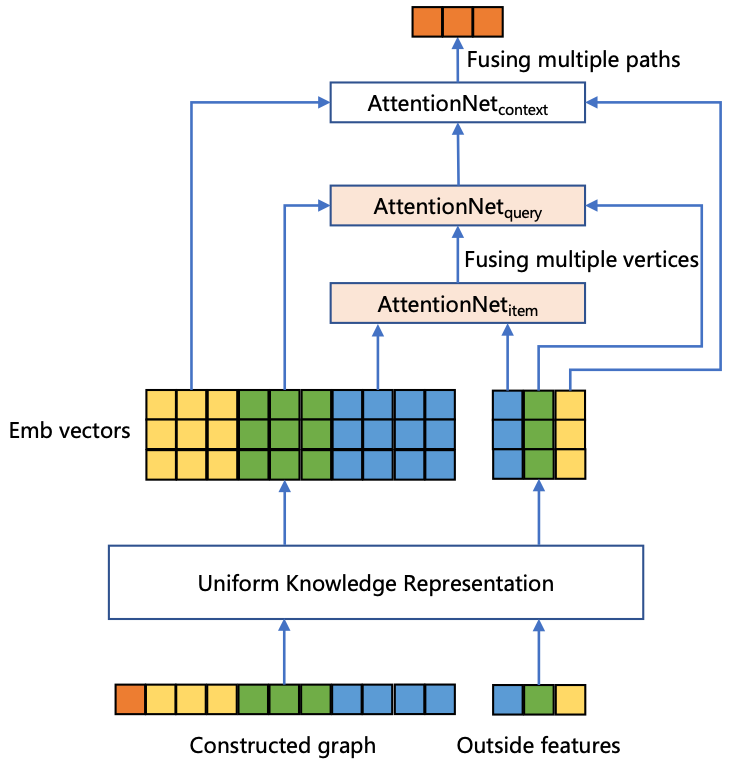}
}\hspace{0.5mm}
\subfigure[Prediction Model]{
\centering
\includegraphics[scale=0.35]{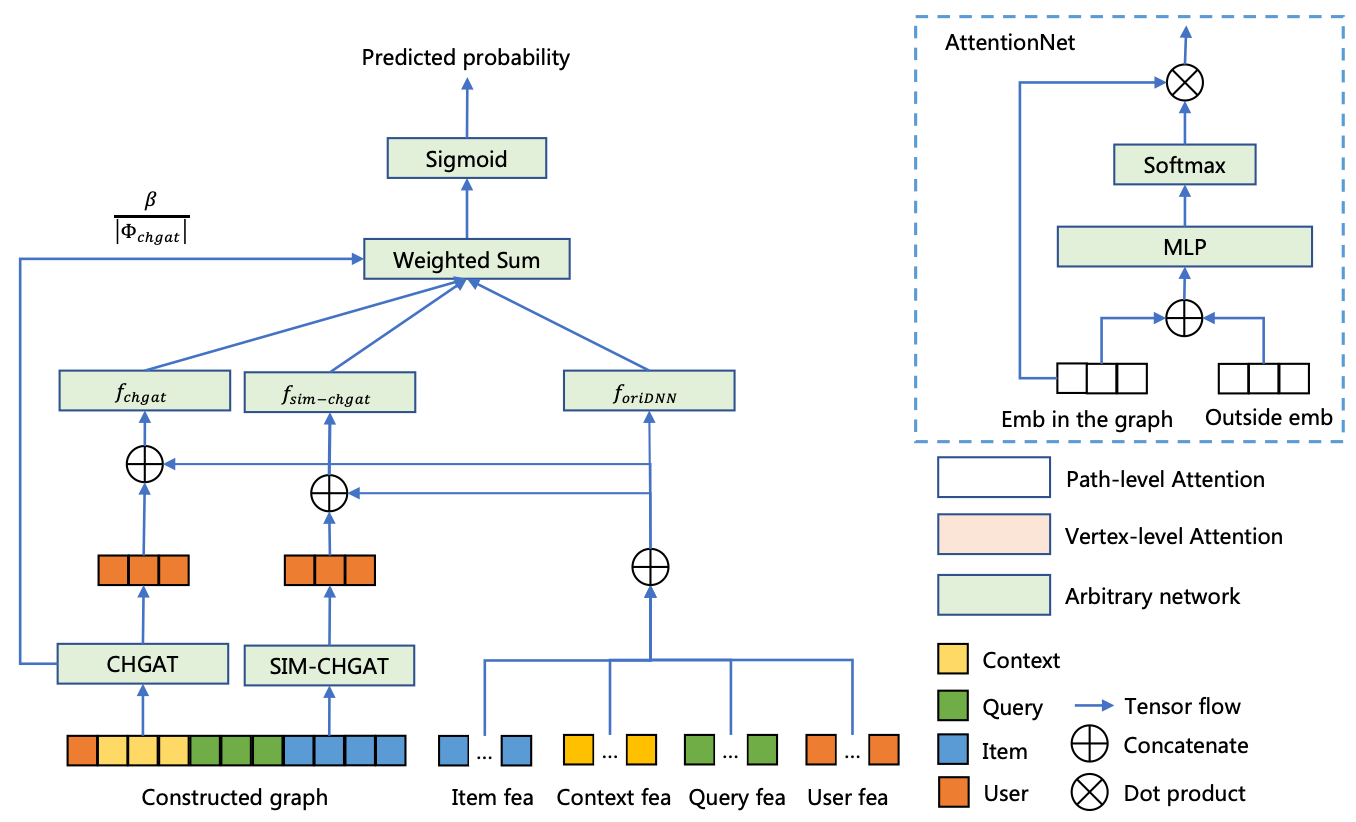}
}\hspace{0.5mm}
\centering
\setlength\abovecaptionskip{-0.1cm}
\setlength\belowcaptionskip{-0.1cm}
\caption{The whole pipeline of our model:(a)Graph Construction,(b)Context-aware Heterogeneous Graph Attention Network,(c)The Complete Prediction Model}
\label{model_framework}
\vspace{-1em}
\end{figure*}

Based on the above analysis, the existing methods cannot naturally solve the above two challenges. Therefore, we propose the \textbf{C}ontext-aware \textbf{H}eterogeneous \textbf{G}raph \textbf{A}ttention ne\textbf{T}work(\textbf{CHGAT}) and introduce the whole pipeline of our model in this paper, which is described in Fig.(\ref{model_framework}). Specifically, the user's own behavior and similar crowds' behavior are firstly utilized to construct heterogeneous graphs for the current user. Then, the uniform knowledge representation is proposed to perform a unified semantic mapping and transformation both on graph vertices and contextual features in the ranking system. Moreover, we introduce a context-aware multi-level attention mechanism to dynamically aggregate graphs by fusing real-time context information. And the detailed aggregation process of our constructed heterogeneous graph is then formulated. Through the multi-level attention mechanism proposed in our research, the heterogeneous behavior graph is able to aggregate the potential user interest according to the current scene to the great extent, which increases the generalization ability of user interest modeling in different scenarios. In general, the main contributions of this paper are as follows:

\begin{itemize}

\item We introduce meta-path based heterogeneous graphs to describe the heterogeneous behavior of users in different contexts, which contains search scenarios and locations. Considering the fact that the limited scale of the user's own behavior may affect the performance of the model, the behavior of similar crowds is innovatively introduced to connect with the pre-constructed behavior graph. In order to capture the characteristics of similar people, the portrait feature of users is chosen as the context feature of similar crowds based behavior graph.

\item We raise a Unified Knowledge Representation(UKR) method based on the knowledge graph to express uniformly the vertices and context features of heterogeneous graphs at the semantic level. UKR can not only provide basic representations for vertices on meta-paths in the process of aggregating user interest but also provide generalized semantic information for external real-time context features. The vertex representation designed in this way reduces the overall parameter amount of the model and adds semantic information to the cold start scenario as well.
\item We propose a heterogeneous graph attention model that is sensitive to the real-time context. Different from the previous approaches, we create a novel multi-level attention mechanism that can be calculated during the graph generation process. Specifically, multiple type-specific vertex-level attention network select vertices that are more related to the current context. The path-level attention network is able to select meta-path based on the relevance of context information at the semantic level. Applying the multi-level attention mechanism, the model proposed in our research is able to aggregate user interest expressions related to specific context scenarios to increase the prediction performance of the ranking system.
\item We collect the real data of the Koubei platform to verify the actual performance of the proposed model. Experimental results show that the proposed model achieves the best performance. In order to verify the auctual application value of CHGAT, we conducted several A/B compare groups, and accumulated millions of browsing data for confident statistical results. The online experiment results prove that CHGAT can well capture contextual information for dynamic aggregation of heterogeneous graphs and return better-matched prediction results.

\end{itemize}

\section{Preliminary}

In this section, we begin by introducing the formulation of the problem that we want to solve. Then, a detailed description of definitions and examples of some key concepts in heterogeneous graphs is given.

\vspace{-0.3cm}
\subsection{Problem Formulaion}

\begin{dfn}[User behavior prediction for the ranking sysytem]
Given a set $<\mathcal{U},\mathcal{Q},\mathcal{I},\mathcal{C},\mathcal{K}>$, where $\mathcal{U}=\{u_1,\cdots,u_P\}$ stands for the set of $P$ users, $\mathcal{Q}=\{q_1,\cdots,q_R\}$ denotes the set of $R$ queries searched by users, $\mathcal{I}=\{i_1,\cdots,i_M\}$ represents the set of $M$ candidate items that remains to be predicted, $\mathcal{C}=\{c_1,\cdots,c_N\}$ stands for the set of $N$ different contexts, and $\mathcal{K}=\{k_1,\cdots,k_S\}$ denotes the set of $S$ uniform knowledge representations. Generally, the overall goal of predicting user behavior for the ranking system in the location-based search(LBS) system is to return more matching items to the user, when they search in specific contexts.
\end{dfn}

In the LBS system of the local consumer service platform, when a user $u\in \mathcal{U}$ searches for a query $q\in \mathcal{Q}$, the ranking algorithm needs to match the user $u$ with the most interesting store $i\in \mathcal{I}$ utilizing the information of search context $c\in \mathcal{C}$, which composed of the real-time location and time of $u$. For example, the user would like to go to a nearer store for a quick lunch at noon on weekdays, and may wish to visit the famous restaurant in the city center to taste unique cuisines on weekend evenings. It can be seen from the above example that the user's interest dynamically changes according to the contextual information. And the ranking algorithm needs to predict a higher score for the candidate item in which the user is more interested in a specific context.

\vspace{-0.3cm}
\subsection{Heterogeneous Graph}\label{2}

In order to generate dynamic interest that changes with the contexts, we propose to utilize heterogeneous interest network as the basic data structure to describe the user behavior, which is also known as a heterogeneous graph $\mathcal{G}=(\mathcal{V},\mathcal{E})$. The heterogeneous graph consists of a vertex set $\mathcal{V}$ and an edge set $\mathcal{E}$. And the whole graph $\mathcal{G}$ can also be associated with a vertex type mapping function $\psi:\mathcal{V}\xrightarrow{}\mathcal{A}$ and a edge type mapping function $\varphi:\mathcal{E}\xrightarrow{}\mathcal{B}$. In the definition of heterogeneous graph, $\mathcal{A}$ and $\mathcal{B}$ represent different types of nodes and edges, respectively, and it should satisfy $\left|\mathcal{A}\right|+\left|\mathcal{B}\right|>2$.

\begin{dfn}[Meta-path]
\vspace{-0.1cm}
Meta-path $\phi$ is defined as the assembly method of different types of vertices and edges in the heterogeneous graph $\mathcal{G}$, which also be considered to include the sequence relationship of the order of vertices.
\end{dfn}

\vspace{-0.1cm}
 Considering that the meta-path represents the semantic meaning of the path formed by relations between $\mathcal{V}$ and $\mathcal{E}$, in this paper, the natural behavior sequences of users are particularly employed to generate the meta-path $\phi$. Therefore, we propose several meta-paths originating from the heterogeneous behavior link, which consists of four types of vertices including $User(U)$, $Item(I)$, $Query(Q)$,$Context(C)$, and their rich interactions. For example, $User-Context-Query-Item(UCQI)$ indicates that the user enters a query in a certain context and then interacts with several items returned by the search engine. And $User-Context-Item(UCI)$ indicates that a user directly interacts with multiple items under a certain context, where the form of interactions include clicking, purchasing, and adding to the shopping cart, etc. Given the premise of the meta-path $\phi$, further analysis of its semantics information requires the definition of neighbor vertices along the meta-path.


\begin{dfn}[Meta-path based Neighbor Vertices]
Meta-path based neighbor vertices $\mathcal{N}_{v}^\phi$ is defined as the set of neighbor vertices of a vertex $v$ on the meta-path $\phi$. 
\end{dfn}



As shown in the Fig.(\ref{model_framework}), for example, in the meta-path $\phi_{UCQI}$, we can get the following neighbor nodes, $\mathcal{N}_{u}^{\phi_{UCQI}}=\{c_1,c_2\}$ represents that the user $u$ has historical behaviors under context $c_1$ and context $c_2$. $\mathcal{N}_{c_1}^{\phi_{UCQI}}=\{q_1,q_2\}$ denotes that user has searched query $q_1$ and query $q_2$ under the context $c_1$, and $\mathcal{N}_{q_1}^{\phi_{UCQI}}=\{i_1,i_2\}$ and represents that the user interacts with item $i_1$ and item $i_2$ after searching for $q_1$.

\section{The Proposed Model}


In this section, we propose a novel supervised graph neural network, which named context-aware heterogeneous graph attention network(CHGAT). For each step of the whole pipeline of our model, we introduce its specific technical details from the background to the target. In the end, we design and assemble a prediction network and define the overall loss function.


\subsection{Overview}
The basic idea of the proposed CHGAT is to design a graph neural network to capture the context-aware potential interest in heterogeneous behaviors and to provide broadened semantic representations for users. As shown in Fig.(\ref{model_framework}), firstly, to prepare the data required by CHGAT, we employ a variety of user behaviors to construct heterogeneous graphs and define contextual features corresponding to different meta-paths. After building the heterogeneous graph, we introduce a unified knowledge representation method that assembles multiple knowledge units to provide unified transformations for different types of vertices in the heterogeneous graph, which can greatly reduce the overall amount of model parameters and represent user interests with clearer semantic information. Moreover, a variety of attention networks that are sensitive to external contextual scenarios are designed, and the calculated attention coefficients are utilized to dynamically aggregate vertices from the same layer to the upper layers in heterogeneous graphs. After the above procedures, the user embedding vector aggregated in the heterogeneous graph is sent to the subsequent deep network for predicting the probability of clicking the candidate $i$ when the user searches for query $q$.

\vspace{-1em}

\subsection{Graph Construction}
\label{graph_construction_section}

\paragraph{Heterogeneous graph constructed from the self-behavior}
For the graph representation method, the information that the model can aggregate is closely related to the way that the graph is constructed. Here, we first apply the historical self behaviors of users to construct the heterogeneous graph according to the two meta-paths of $\phi_{UCQI}$ and $\phi_{UCI}$, and only retain the historical behavior related edges during the construction. And considering the requirements of the LBS system, we set the location where the user's historical behavior occurred as the context vertex $C$, and merge the similar behaviors that occurred in the same context into the same meta-path. Meanwhile, the central vertex in the graph is defined as the root vertex in the meta-path, which is conducive to the subsequent graph aggregation process to maximize the retention of hidden interest information in the original behavior sequence. It is worth noting that our graph is constructed from users, so we set the user as the root vertex $v_r$ of the constructed graph.

\paragraph{Heterogeneous graph transferred from similar crowds} However, when the user has a small number of historical behaviors, the scale of the constructed heterogeneous graph is limited, which in turn will affect the prediction effect of the whole model. In response to this problem, we additionally introduce a heterogeneous graph constructed based on heterogeneous behaviors of similar crowds. Considering the spatial characteristics of the LBS scenario, we treat users within a certain distance from the current user's search location as similar crowds and utilize their most recent interaction behavior as a historical behavior database of similar crowds to build the graph. At the same time, we set the feature of the user portrait as the context vertex $C$ to facilitate the selection of the user expression that is most similar to the current user from the behavior of other people during subsequent graph aggregation.

After establishing the user's own heterogeneous graph and the heterogeneous graph of similar crowds, we connect them to the shared root vertex, which is the user to be represented in the current ranking system. At this time, the complete heterogeneous graph contains four types of meta-paths, which names $\phi_{UCQI_{self}}$, $\phi_{UCI_{self}}$, $\phi_{UCQI_{sim}}$, and $\phi_{UCI_{sim}}$. And before the subsequent aggregation, all the vertices in the graph used their original ID for recording.


\begin{figure}[] 
\centering 
\includegraphics[width=0.25\textwidth]{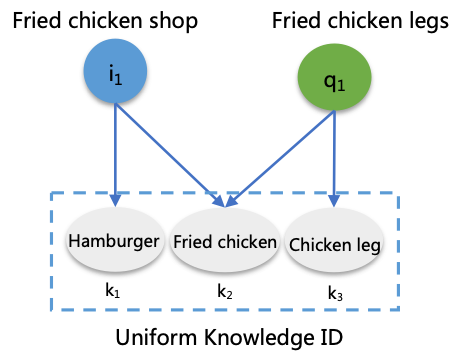} %
\setlength{\abovecaptionskip}{0cm}\caption{An example of the uniform knowledge representation} 
\label{UKR} 
\vspace{-2em}
\end{figure}

\subsection{Uniform knowledge representation}
\label{Uniform knowledge representation}
One of the key challenges of applying heterogeneous graphs in ranking algorithms is to comprehend the heterogeneous vertices in the graph. But if directly employing the original ID of the vertex for embedding or use its pre-trained embedding as the expression of the vertices in the graph, there may exist following two problems: 

\begin{itemize}
\item With tens of millions of vertices in the heterogeneous graph, a separate embedding expression for each node in the graph will cause the model to have a tremendous amount of parameters, which in turn affect the availability and time-consuming performance of the model.
\item In the heterogeneous graph, the vertex to be aggregated and its neighbor vertices in the same meta-path are often heterogeneous, which indicates that the embedding of original vertices is not very semantically related. It also denotes that during the process of graph aggregation, there would be conflicts of the heterogeneous information between neighbor vertices and the vertex to be aggregated, which eventually increases the difficulty of the converging of model parameters.
\end{itemize}

In order to avoid the impact of the above problems on the performance of the model, we introduce the uniform knowledge representation(UKR) to comprehend vertices in the heterogeneous behavior graph. The conversion range of the unified knowledge expression includes all types of vertices in the meta-path except the root vertex. For example, in our research, the main vertices are query, item, and search scene lies both in constructed meta-paths and the input of the ranking system.

Specifically, we first utilize the knowledge graph\cite{wang2014knowledge} to extract key knowledge about the store in our platform, which includes the primary business category of the store, the main tag in the title of the store, and the name of commodities in the store. Next, the search comprehension engine is used to predict and retain the key knowledge of search intent, search category, text entity, and so on. And at the aspect of comprehending search scenarios, locations, nearby high-frequency shops, weather, and time, are selected as the knowledge representation.

It is worth noting that the knowledge representation extracted from the original vertex is first retained in the form of several texts, and the knowledge expressed in the same space can be extracted from different heterogeneous materials. As shown in the Fig.(\ref{UKR}), a fried chicken shop $i_1$ contains the main knowledge of hamburger $k_1$, fried chicken $k_2$. And the query $q_1$ search for fried chicken legs also contains fried chicken $k_2$. At this time, the knowledge unit $k_2$ of representing fried chicken for these two heterogeneous vertices is the same.

 Then we apply the conversion relationship between all heterogeneous materials and knowledge representations to construct a key-value pair knowledge dictionary $\mathcal{K}$, the key in the dictionary is the original id of vertices in the graph, and value is multiple converted knowledge units. Taking Fig.(\ref{UKR}) as an example, $\mathcal{K}=\{i_{1}:\{k_{1},k_{2}\},q_{1}:\{k_{2},k_{3}\}\}$ represents the knowledge dictionary.

 After mapping the original ID of the vertex in the heterogeneous graph to the knowledge unit $k$, a function $m:\mathcal{K}\rightarrow\mathbb{R}^d$ is designed to respectively map the $k$ after the one-hot transformation to the $d$-dimensional embedding vector $e$. Note that each original vertex in the heterogeneous graph is composed of multiple knowledge units, so it is necessary to introduce an knowledge aggregation function $g_k$ to fuse multiple knowledge embedding vectors as the embedding vector of the current vertex. In the above example, the embedding of the fried chicken shop $i_1$ and the embedding of the query $q_1$ of fried chicken legd are represented as following
\setlength\abovedisplayskip{1.5pt}
\setlength\belowdisplayskip{1.5pt}
\begin{equation}
\begin{split}
e_{i_1}&={g_k}(e_{k_1},e_{k_2})={g_k}(m({k_1}),m({k_2}))\\
e_{q_1}&={g_k}(e_{k_2},e_{k_3})={g_k}(m({k_2}),m({k_3}))
 \end{split}
\end{equation}
where the knowledge aggregation function $g_k$, which can be assembled by the neural network, weighted sum, and other methods, determines the current vertex knowledge assembly method and knowledge focus. In order to incorporate more information, an element-wise average function is selected as the function $g_k$ in this paper.

 By applying the above approach, it is convenient for us to uniformly express large-scale original materials in the form of assembly knowledge units, which greatly reduces the scale of model parameters and solve the problem of the semantic gap between adjacent vertices to a certain extent. Meanwhile, for a new vertex in a graph, we can also utilize the trained knowledge embeddings to quickly comprehend it. It is worth noting that the embedding parameters here are part of the model and are also trained with the CHGAT main model. Referring to the Table.(\ref{auc_table}), this design further improves the overall prediction accuracy of the model. 


\subsection{Context-aware Heterogeneous Graph Attention Network}
\label{CHGAT section}


In the process of the whole pipeline, another key challenge of utilizing the meta-path guided heterogeneous graphs is the selection and aggregation of vertices in the heterogeneous graph. Different from other graph aggregation methods that utilize the relationship between neighbor vertices in the complete graph\cite{velivckovic2017graph}, we propose a new multi-level attention based graph aggregation mechanism, which fuses the outside contextual features to aggregate vertices in the constructed graph. Especially, considering the fact that vertices on the meta-path are divided into the root vertex and its neighbor vertices, we introduce two design schemes of attention mechanism for the topological characteristics of the heterogeneous graph. It is also worth noting that the external real-time outside features utilized below include the real-time search scenario of the user, the current search query in the ranking system, and the candidate item to be predicted. For the convenience of expression, we call such features as outside vertices $v_o$.


We first introduce a novel vertex-level attention mechanism that can learn the importance of different vertices in the current contextual scene and gradually aggregate the meaningful information in the vertices along the meta-path. Given a vertices pair $(v_i,v_j),{v_j}\in \mathcal{N}_{v_i}^\phi$, the vertex-level attention aims to get the attention weight $\alpha_{v_i,v_j}^\phi$ that determines the importance of vertice $v_i$ to vertice $v_j$. Different from methods such as HAN\cite{han2019} which directly utilize the vertex pair $(v_i,v_j)$ to calculate the weight, we introduce the real-time outside vertex $v_o$ of the same type with vertex $v_j$ to calculate the attention coefficient:
\setlength\abovedisplayskip{1.5pt}
\setlength\belowdisplayskip{1.5pt}
\begin{equation}
\alpha_{v_i,v_j}^\phi=Attention_{\psi(v_j)}(v_j,v_o)
\end{equation}
where $v_i$ stands for the vertex to be aggregated, and $Attention_{\psi(v_j)}$ refers to a specially designed vertex-level attention network for the type of vertex $v_j$, and $\phi\in \Phi$ is one of the meta-paths in the heterogeneous graph.

In order to better construct the correlation between external vertex and vertices in the graph, the uniform knowledge representation is applied to map and transform these vertices before inputted to the attention network. And the detailed formulation of the attention weight can be calculated as follows:
\begin{equation}\label{1}
\alpha_{v_i,v_j}^\phi=\frac{exp(\sigma(MLP_{\psi(v_j)}(e_{v_j},e_{v_o})))}{\sum_{v\in\mathcal{N}_{v_i}^\phi}exp(\sigma(MLP_{\psi(v)}(e_v,e_{v_o})))}
\end{equation}
where $\mathcal{N}_{v_i}^{\phi}$ stands for the meta-path based neighbor upstream vertices of $v_i$, and $MLP_{\psi(v)}$ denotes a multilayer perceptron for specific type of the current vertex, $\sigma$ represents the activation function. Then, introducing the vertex-level aggregation function $g_r$, the aggregated representation of the vertex $v_i$ can be obtained as:
\setlength\abovedisplayskip{1.5pt}
\setlength\belowdisplayskip{1.5pt}
\begin{equation}
\begin{split}
e_{v_i}^\phi&=g_{r}(\alpha_{v_i,v}^\phi,e_v), {v\in \mathcal{N}_{v_i}^\phi}\\
\end{split}
\end{equation}
where $e_{v_i}^\phi$ is the aggregated embedding vector for vertex $v_i$ along the meta-path $\phi$, and the aggregation function $g_{r}$ can be any other approaches that can incorporate the attention weight coefficients, such as attention-based LSTM\cite{wang2016attention}, etc.


Furthermore, after utilizing the vertex-level attention to obtain the correlation weight of each vertex and aggregate the representation along the meta-path $\phi$, the number of aggregated representations $e^\phi, \phi\in \Phi$ that connected with the root vertex $v_r$ is consistent with the number of pre-defined meta-paths in the graph, which is $\left|\Phi\right|_{\mathcal{G}}$. And in this paper, as described in Subection\ref{graph_construction_section}, each meta-path $\phi$ has a unique context feature $c_\phi$. Therefore the path representation $e_{v_r}^\phi$ equals to the embedding vector aggregated to the vertices of the context $e_{c}^\phi$ along the meta-path $\phi$. In order to obtain the unique representation of the root vertex, we then propose the path-level attention to fuse multiple meta-path representations in the graph, which can be defined as:
\setlength\abovedisplayskip{1.5pt}
\setlength\belowdisplayskip{1.5pt}
\begin{equation}
\alpha_{v_r}^\phi=Attention_{\psi(c_\phi)}(c_\phi,c_o)
\end{equation}

where $\alpha_{v_r}^\phi$ is the attention weight of meta-path $\phi$ for the root vertex $v_r$, and $c_o$ is the outside context feature, and $\psi(c_\phi)$ is the type of the context feature $c_\phi$. 
The goal of the path-level attention is to use the information of external context characteristics to filter the most relevant path of the heterogeneous graph and choose the representation of the meta-path that is most similar to the current context. In order to achieve this target, and considering that users’ interests do not explicitly include contextual features, we use the self-representation $e_{c_{\phi}}$ of the context feature $c_\phi$ to calculate the weight coefficient, and utilize the aggregated representation $e_{v_r}^{\phi}$ from the meta-path $\phi$ as the path representation together with the weight coefficient $\alpha_{v_r}^\phi$ to get the final representation of the root vertex $v_r$. Hence, the detailed formula of the path-level attention network is as follows:
\setlength\abovedisplayskip{1.5pt}
\setlength\belowdisplayskip{1.5pt}
\begin{equation}\label{3}
\alpha_{v_r}^\phi=\frac{exp(\sigma(MLP_{\psi(c_\phi)}(e_{c_\phi},e_{c_o})))}{\sum_{c_{\phi'}\in\mathcal{N}_{v_r}}exp(\sigma(MLP_{\psi(c_\phi')}(e_{c_{\phi'}},e_{c_o})))}
\end{equation}


where $c_{\phi'}$ are context vertices that belong to the neighbor of root vertex $v_r$. Therefore, in the design concept of the path-level attention, the representation of the root vertex can be calculated as the aggregation from all meta-paths, where we introduce $g_p$ as the aggregate function:
\setlength\abovedisplayskip{1.5pt}
\setlength\belowdisplayskip{1.5pt}
\begin{equation}\label{4}
e_{v_r}=g_p(\alpha_{v_r}^\phi,e_{v_r}^\phi),{\phi\in\Phi}
\end{equation}
where $\Phi$ denotes all meta-paths in the heterogeneous graph, and $e_{v_r}^\phi$ refers to the representation of different meta-paths of the root vertex $v_r$, and $e_{v_r}$ stands for the final representation of the root vertex. Next, we apply the above two attention mechanisms to the aggregation process of heterogeneous behavior graphs constructed in Subsection\ref{graph_construction_section}.


\paragraph{Aggregation process in $\phi_{UCQI_{self}}$ and $\phi_{UCI_{self}}$ based graph} 
At this time, the behavior in the heterogeneous graph based on $\phi_{UCQI_{self}}$ and $\phi_{UCI_{self}}$ comes from the user itself and records the user's active interaction behavior in different contexts. Considering that the historical behavior structure in this type of meta-path is similar to the current prediction scene, so the real-time input query and the item to be predicted in the ranking algorithm are utilized as the outside vertex $v_o$ of the vertex-level attention. For example, as shown in Fig.(\ref{model_framework}), when the information of vertex $v_{i_1}$ is aggregated to vertex $v_{q_1}$ along the meta-path $\phi_{UCQI_{self}}$, the outside item $v_{i_o}$ to be predicted and the current vertex $v_{i_1}$ are used as the input of the attention network $Attention_{item}$, which can be detailed as:
\setlength\abovedisplayskip{1.5pt}
\setlength\belowdisplayskip{1.5pt}
\begin{equation}
\label{vertex-example}
\alpha_{q_1,i_1}^{UCQI_{self}}=\frac{exp(\sigma(MLP_{item}(e_{i_1},e_{i_o})))}{\sum_{i\in\mathcal{N}_{q_1}^{UCQI_{self}}}exp(\sigma(MLP_{item}(e_i,e_{i_o})))}
\end{equation}
After multi-layer vertex-level aggregation along the meta-path, the representation $e_u^{\phi}$ of each meta-path $\phi$ can be obtained. Considering the characteristics of the LBS scenario, we directly apply the search location as the context feature of each meta-path, and at this time $c_\phi$ represents the recorded user search location for each meta-path $\phi$ in the historical behavior graph. Then the importance of meta-path $\phi$ to the current user $u$ can be obtained by the Eq.(\ref{3}). And using the weights generated by path-level attention to fuse the representation of multiple meta-paths, the user representation $e_u^{chgat}$ of the $\phi_{UCQI_{self}}$ and $\phi_{UCI_{self}}$ based graph can be obtained.

\paragraph{Aggregation process in $\phi_{UCQI_{sim}}$ and $\phi_{UCI_{sim}}$ based graph}

For heterogeneous graphs constructed from other similar crowds behaviors, which contains the $\phi_{UCQI_{sim}}$ and $\phi_{UCI_{sim}}$, the aggregation at the vertex-level is consistent with the Eq.(\ref{vertex-example}), and the basic representations of each path can be obtained, which is also the representation of similar crowds. In order to select the user representation most similar to the current user among similar crowds, we apply the basic portrait feature of users as the context feature $c$ in Eq.(\ref{3}), for example, the path-level attention weight of the meta-path $\phi$ to the present user $u$ can be obtained as:
\setlength\abovedisplayskip{1.5pt}
\setlength\belowdisplayskip{1.5pt}
\begin{equation}
\alpha_{u}^{\phi}=\frac{exp(\sigma(MLP_{user}(e_{c_{\phi}},e_{c_u})))}{\sum_{c_{\phi'}\in\mathcal{N}_{u}}exp(\sigma(MLP_{user}(e_{c_{\phi'}},e_{c_u})))}
\end{equation}
where $e_{c_u}$ is the basic portrait feature of the present user $u$. Same as the Eq.(\ref{4}), applying the weighted sum as the function to aggregate embeddings of meta-paths:
\setlength\abovedisplayskip{-2pt}
\setlength\belowdisplayskip{-2pt}
\begin{equation}\label{weighted_sum}
e_{u}=\sum_{\phi\in \Phi}(\alpha_{u}^\phi,e_{u}^\phi)
\end{equation}
where $e_{u}^\phi$ is the aggregated embedding vector for meta-path $\phi$, $e_{u}$ is the final representation originated from similar crowds, which is also named $e_u^{sim-chgat}$ in our paper.

In conclusion, the focus of the vertex-level attention network is to capture the correlation between the vertices to be aggregated and outside vertex of the same type and to select the vertex that is most similar to the current outside feature as the vertex-level aggregation information passed down. From the perspective of the user interest, the vertex-level attention is able to choose the item that most similar to the current outside item from the historical behavior as the basic representation of interest. Moreover, the purpose of the path-level attention network is to predict the correlation between path-level contextual information and current contextual attributes in the ranking system, and then choose the meta-path that is most similar to the current outside context feature as the primary representation of the root vertex in the graph.

\vspace{-1em}

\subsection{The Loss Function}


After obtaining $e_{u}^{chgat}$ and $e_{u}^{sim-chgat}$, we introduce two networks $f_{chgat}(\cdot)$ and $f_{sim-chgat}(\cdot)$ to separately obtain their logits, which are then accumulated with the logits of the original network as the final output of the network. Therefore, the probability ${\widehat y}_{u,q_o,i_o,c_o}$ denotes that user $u$ searches for the query $q_o$ and clicks the candidate $i_o$ in a certain outside search context $c_o$, which is predicted by our proposed CHGAT model, as Fig.(\ref{model_framework}) shows, is established as:
\begin{equation}
\label{concat_embedding}
\begin{split}
{\widehat y}_{u,q_o,i_o,c_o}&=sigmoid(f_{chgat}(e_u^{chgat})\vert\vert f_{attri}(u,q_o,i_o,c_o)  \\&+ \frac\beta{\left|\Phi_{chgat}\right|}\cdot f_{sim-chgat}(e_u^{sim-chgat})\vert\vert f_{attri}(u,q_o,i_o,c_o) \\&+ f_{oriDNN}(f_{attri}(u,q_o,i_o,c_o)))
\end{split}
\end{equation}
where $\{u,q_o,i_o,c_o\}$ denotes the basic element of the search prediction system, $f(\cdot)$ can be any form of deep neural network, $f_{attri}(\cdot)$ represents multiple types of attributes, such as the features described for $u$, $q_o$, $i_o$, and $c_o$. It is worth noting that $\frac\beta{\left|\Phi_{chgat}\right|}$ measures the number of meta-paths in the self-behavior graph, especially when the user's original heterogeneous behavior graph is small, it will improve the influence of similar crowds to our model. And $\beta$ is a hyperparameter for the SIM-CHGAT part.

The loss function $\mathcal L$ defines the discrepancy between the predicted probability ${\widehat y}_{u,q_o,i_o,c_o}$ of the model and the true probability ${y}_{u,q_o,i_o,c_o}$. Here we utilize the cross entropy function as the loss function:
\begin{equation}
\begin{split}
\mathcal L =&-\sum_{(u,q_o,i_o,c_o)}\lbrack y_{(u,q_o,i_o,c_o)}\log({\widehat y}_{(u,q_o,i_o,c_o)})\\&+(1-y_{(u,q_o,i_o,c_o)})\log(1-{\widehat y}_{(u,q_o,i_o,c_o)})\rbrack \\&+ \lambda R(\theta)
\end{split}
\end{equation}
where $R(\theta)$ denotes the regularization function for parameters of the whole model, and $\lambda$ is the hyperparameter for the regularization part. And in the model training process, the Adam optimizer is utilized to minimize the loss function\cite{kingma2014adam}.

\section{Experiments}

In this section, we first employ experiments on datasets of the Koubei platform to compare the proposed CHGAT with up-to-date state-of-the-art methods. Next, the sensitivity of the model performance to multiple hyperparameters is also verified, including the number of uniform knowledge units in the UKR part and the $\beta$ in Eq.(\ref{concat_embedding}). Moreover, we also examine and obtain the A/B results of the model on the actual online local e-commerce system.

\begin{table}[]
\scalebox{0.7}{
\begin{tabular}{c|c|c|c|c}
\hline
\hline
Datasets    & \# Sample & \# Positive sample  & \# $E(\left|\Phi\right|_{chgat})$ & \# $E(\left|\mathcal{V}\right|_{chgat})$ \\ \hline
Full-week Train      &      $3.26\times10^8$      &        $6.27\times10^7$    &        $5.35$       &         $32.41$          \\ 
Full-day Test &     $7.24\times10^6$        &            $1.38\times10^6$         & $5.17$                &           $30.94$         \\ 
Full-week Test &     $8.16\times10^6$       &            $1.51\times10^6$         & $5.41$               &        $33.18$            \\ 
Full-week Hard Test  &      $2.29\times10^6$      &           $1.78\times10^5$          &   $0.85$            &    $4.71$                  \\ 
\hline
\hline
\end{tabular}}
\caption{Basic Description of Datasets}
\label{Basic_Description_of_Datasets}
\vspace{-3em}
\end{table}

\subsection{Datasets}
We collect real online data from the leading local consumer service platform in China, the Koubei app\footnote{https://www.koubei.com}. Specifically, the offline dataset covers a consecutive week's true behavior, which is taken measures such as negative sampling and noise filtering before further employed. Then the dataset can be described from the following multiple perspectives:

\begin{itemize}
\item From the perspective of feature generation, we have constructed the attribute features, statistical features, sequence features, and category features of users, queries, scenes, and shops. These basic features constitute the attribute feature parts in the Eq.(\ref{concat_embedding}).

\item From the perspective of the user's own heterogeneous graph, we utilize the user's historical behavior in the past 30 days to construct a heterogeneous behavior graph, which mainly contains the meta-path $\phi_{UCQI_{self}}$ and the meta-path $\phi_{UCI_{self}}$.

\item From the perspective of the heterogeneous graph of similar crowds, we make use of search behaviors of people within three kilometers from the user’s current search position to construct a heterogeneous graph. The meta-path in the graph mainly includes $\phi_{UCQI_{sim}}$ and $\phi_{UCI_{sim}}$, where the number of each type of meta-path is limited up to $20$.
\end{itemize}

As shown in the Table.(\ref{Basic_Description_of_Datasets}), we randomly sample a major part of a whole week’s dataset as the full-week training dataset and another small part as the full-week test dataset. And the purpose of doing like this is to reduce the impact of different dates on the sample distribution as much as possible. Moreover, the full-day test dataset utilizes random sampling throughout the day on a certain day after a week to simulate actual prediction scenarios. As for the statistic attributes of the heterogeneous graph in Table.(\ref{Basic_Description_of_Datasets}), $E(\left|\Phi\right|_{chgat})$ refers to the expectation value of the number of meta-paths per sample and $E(\left|\mathcal{V}\right|_{chgat})$ denotes the the expectation value of the number of edges in the graph. It is worth noting that we select the sample of few-behaving users to construct a full-week hard test dataset. The $E(\left|\Phi\right|_{chgat})$ in this dataset is much smaller than other datasets, which aims to verify the effect of the proposed SIM-CHGAT.

\begin{table*}[ht]
\scalebox{1}{
\begin{tabular}{c|c|c|c|c|c|c}
\hline
\hline
\multirow{2}{*}{Method} & \multicolumn{2}{c|}{Full-day} & \multicolumn{2}{c|}{Full-week} & \multicolumn{2}{c}{Full-week hard} \\ \cline{2-7} 
                        & AUC           & NDCG          & AUC           & NDCG           & AUC              & NDCG             \\ \hline
LR           &     $0.7620$          &       $0.5120$        &        $0.7709$       &          $0.5097$      &         $0.7204$         &        $0.4830$          \\ \hline
DNN           &     $0.7763$           &     $0.5266$           &       $0.7811$        &         $0.5241$       &         $0.7362$         &       $0.4975$           \\ \hline
Wide\&Deep(WD)           &    $0.7795$           &     $0.5289$          &    $0.7859$           &      $0.5308$           &     $0.7366$             &        $0.5017$          \\ \hline \hline
DIN-WD                  &     $0.7816$          &   $0.5407$            &     $0.7938$          &        $0.5382$        &           $0.7371$        &     $0.5094$              \\ \hline
Query-DIN-WD                &     $0.7860$       &      $0.5396$         &     $0.7951$        &       $0.5401$         &           $0.7354$       &        $0.5088$          \\ \hline \hline
HAN-WD                   &    $0.7741$           &       $0.5325$        &      $0.7826$         &      $0.5327$           &     $0.7309$             &        $0.5051$          \\ \hline
MEIRec-WD                   &    $0.7807$           &       $0.5364$        &      $0.7919$         &     $0.5412$           &        $0.7317$          &     $0.5092$             \\ \hline \hline
noUKR-CHGAT            &    $0.7871$            &     $0.5390$          &        $0.7942$       &       $0.5327$         &            $0.7345$      &        $0.5063$          \\ \hline
CHGAT                   &      \bm{$0.8018$}        &       $0.5453$        &        $0.8059$       &        $0.5420$        &             $0.7360$     &       $0.5097$           \\ \hline
SIM-CHGAT               &     $0.8006$     &      \bm{$0.5471$}          &       \bm{$0.8120$}        &          \bm{$0.5494$}     &           \bm{$0.7591$}       &       \bm{$0.5215$}           \\ \hline

Improvement &     $2.01\%$     &      $1.18\%$          &       $2.12\%$         &        $1.51\%$      &       $2.97\%$           &           $2.31\%$        \\ 
\hline
\hline
\end{tabular}}
\setlength\belowcaptionskip{0.1cm}
\caption{The ranking metrics of different methods. The best results are indicated in bold, and the last row denotes the improvement of the method proposed in our research compared to the best baseline method}
\label{auc_table}
\vspace{-0.5em}
\end{table*}

\subsection{Baseline Methods and Experimental Settings}

In order to verify the performance of the proposed model, we utilized the latest prediction model in the industry and methods related to our model as baseline methods to create offline compare groups, which can be described as:

\begin{itemize}
\item \textbf{Logistic Regression}\cite{hosmer2013applied} is a basic linear model, which employs statistical features and one-hot features predict the probability of classification tasks.
\item \textbf{Deep Neural Network}\cite{45530} is a neural network with multiple layers, which is able to transform categorical features into embedding vectors.
\item \textbf{Wide\&Deep}\cite{DBLP:journalscorrChengKHSCAACCIA16} combines LR and DNN to balance memory performance and generalization performance, which is choosed as the original model $f_{oriDNN}(\cdot)$ for following models.
\item \textbf{DIN-WD}\cite{zhou2018deep} utilizes the item sequence of the users' past interactions to model interest representation. In our experiments, we combine it with WD to predict the click-through rate.
\item \textbf{Query-DIN-WD} adds a query sequence more than the original DIN-WD model.
\item \textbf{MEIRec-WD}\cite{10.1145/3292500.3330673} builds a heterogeneous graph based on multiple artificial meta-paths, which is selected as the comparison of graph aggregation methods. 
\item \textbf{HAN-WD}\cite{han2019} employs the correlation between neighbor nodes of the heterogeneous graph to aggregate. In our experiments, we treat it as a comparison from the perspective of the graph attention network.
\item \textbf{noUKR-CHGAT} is the proposed model in our research. However, it is short of the uniform knowledge representation part, which is detailed in Section\ref{Uniform knowledge representation}.
\item \textbf{CHGAT} is the proposed model in Section\ref{CHGAT section}, which lacks the embedding aggregated from the similar crowds graph.
\item \textbf{SIM-CHGAT} is the complete version of the model proposed in our research.

\end{itemize}

\begin{figure*}[htbp]
\vspace{-0.8cm}
\centering
\subfigure[Full-day]{
\centering
\includegraphics[scale=0.2]{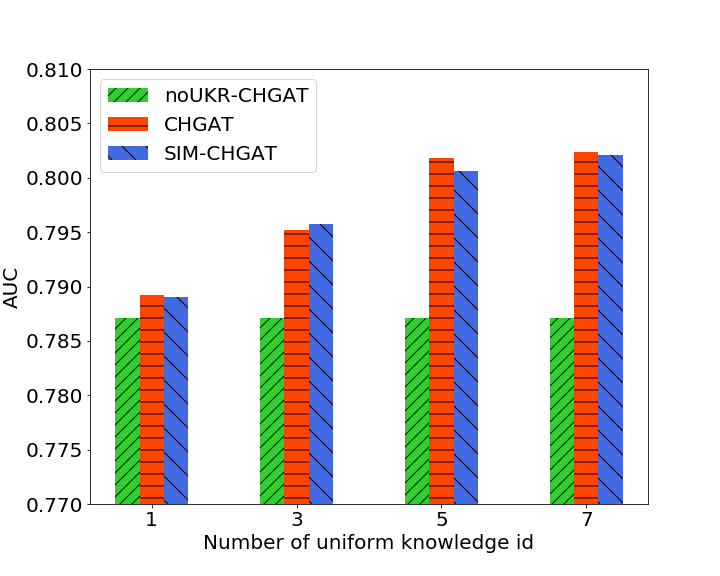}
}\hspace{0.5mm}
\subfigure[Full-week]{
\centering
\includegraphics[scale=0.2]{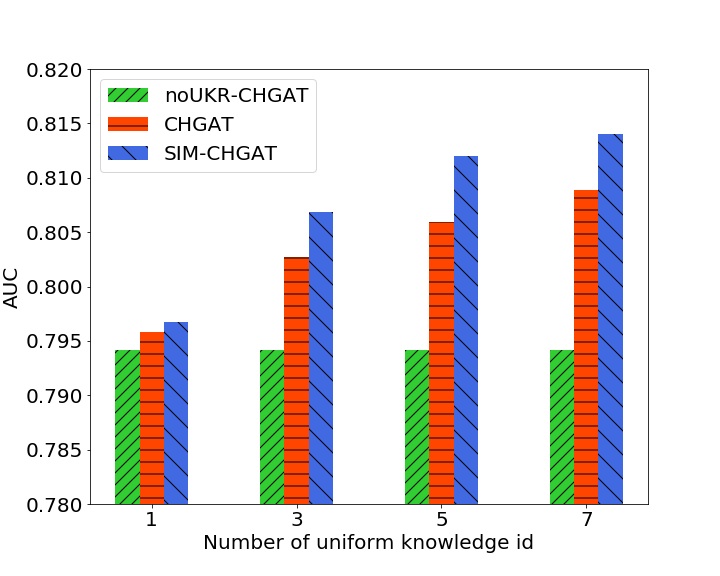}
}\hspace{0.5mm}
\subfigure[Full-week hard]{
\centering
\includegraphics[scale=0.2]{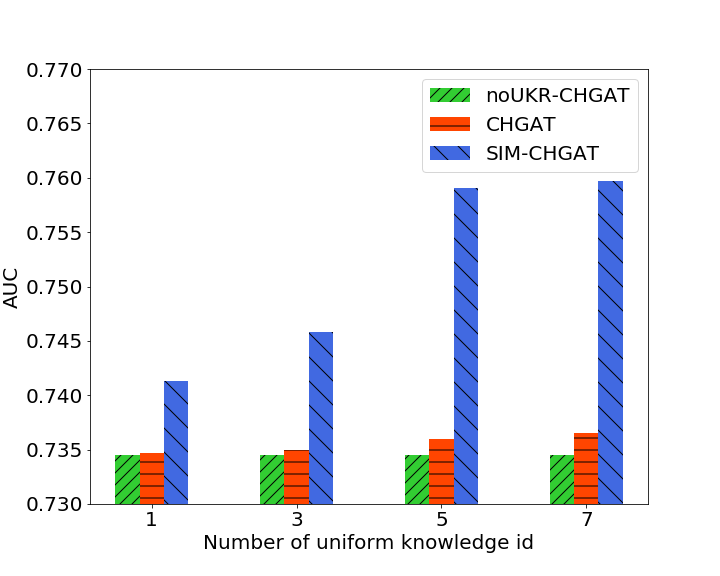}
}\hspace{0.5mm}
\centering
\setlength{\abovecaptionskip}{-0.1cm}
\caption{Parameter sensitivity of the number of uniform knowledge id}
\label{UKR_number}
\vspace{-1em}
\end{figure*}


\begin{figure*}[htbp]
\centering
\subfigure[Full-day]{
\centering
\includegraphics[scale=0.18]{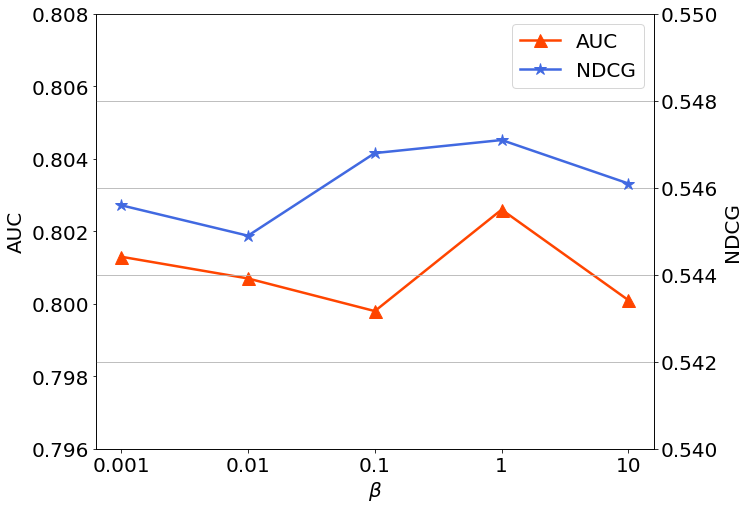}
}\hspace{0.5mm}
\subfigure[Full-week]{
\centering
\includegraphics[scale=0.18]{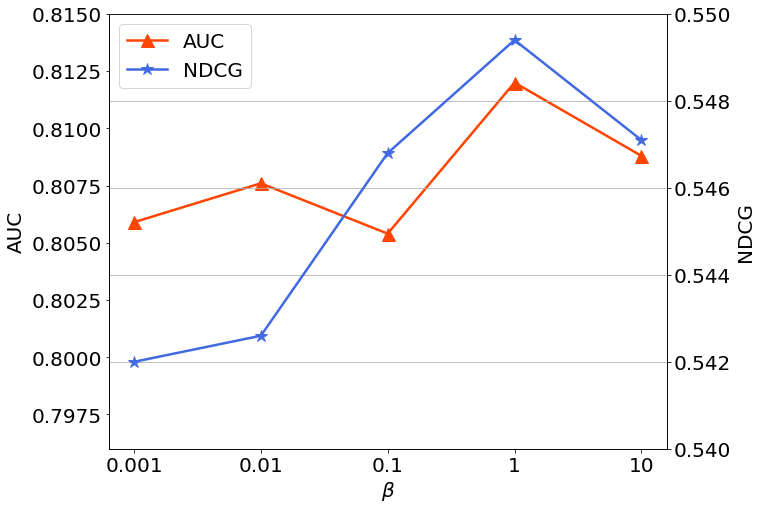}
}\hspace{0.5mm}
\subfigure[Full-week hard]{
\centering
\includegraphics[scale=0.18]{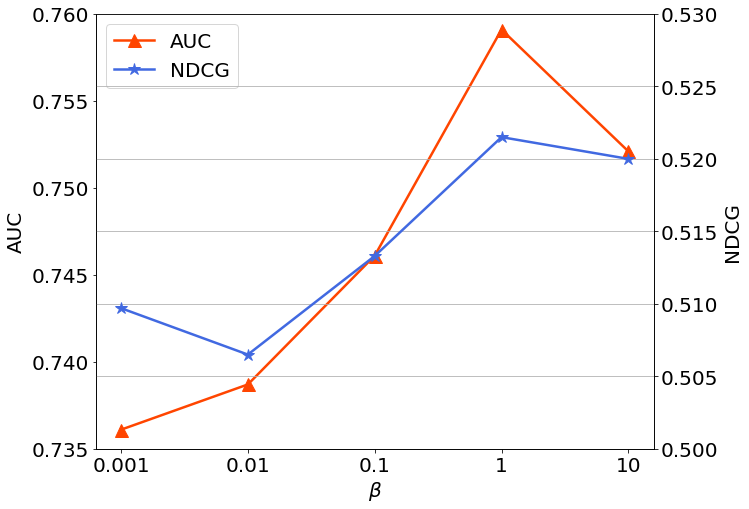}
}\hspace{0.5mm}
\centering
\setlength{\abovecaptionskip}{-0.1cm}
\caption{Parameter sensitivity of the $\beta$ for SIM-CHGAT part}
\label{beta_auc}
\vspace{-1em}
\end{figure*}

For the above-selected methods, the shared data, features, and hyperparameters of the model are all kept consistent to establish corresponding control groups. And the aggregation function $g$ in our model is set to be the weighted sum. And we utilize $Relu$ as the activation function\cite{agarap2018deep} and set the $\lambda$ equals $0.01$. In order to compare the performance of the model on user behavior prediction tasks, we leverage AUC\cite{lobo2008auc} and average NDCG\cite{valizadegan2009learning} as the evaluation indicators.

\subsection{Overall Performance}

In order to evaluate the prediction performance of the proposed model in different scenarios, we verified the performance of the baseline and our algorithm in three test datasets, and the obtained results are shown in the Table.(\ref{auc_table}). According to the detailed data, we now give analyses of experiment results::
\begin{itemize}
\vspace{-0.3em}
\item Under evaluate datasets employed in our paper, performances of CHGAT and SIM-CHGAT are both significantly better than other baseline methods. Specifically, on the Full-day dataset that simulates the actual online prediction scenario, CHGAT improves the AUC by $2.01\%$ compared to QUERY-DIN-WD, indicating that the performance of the method proposed in this article is better than other methods in the actual prediction scenario. With a longer time distribution, the full-week dataset performance of CHGAT and SIM-CHGAT is significantly better than other graph methods or behavior sequence methods, indicating that the proposed CHGAT model is able to describe dynamic user interest which varifies with different contexts, and returns more matching items for users.

\item The noUKR-CHGAT employs graphs that have not been transformed with the unified knowledge representation. As can be seen in Table.(\ref{auc_table}), in the prediction results of all test datasets, noUKR-CHGAT is significantly weaker than CHGAT with UKR whether in the aspect of AUC or NDCG. In addition, the number of knowledge IDs in UKR represents the amount of information contained in the transformed vertices, where the noUKR-CHGAT has no knowledge IDs. As we can see in the Fig.(\ref{UKR_number}), as the number of knowledge id increases, the AUC of CHGAT in multiple test datasets can be significantly improved, indicating that our proposed UKR method is able to effectively establish the semantic relationship between heterogeneous behavior and heterogeneous materials. It is also worth noting that the employ of UKR can reduce the model size from $1.72G$ to $0.59G$, which improves the training convergence effect of the model. And we set the number of knowledge id equals to $5$.

\item We verify the performance of SIM-CHGAT model under multiple test conditions. It can be seen from the right column in Table.(\ref{auc_table}) that in the hard test dataset, the performance of SIM-CHGAT is better than other methods, which proves that the introduction of similar crowd behavior has a better impact on cold-start user behavior prediction. In addition, as shown in the Fig.(\ref{beta_auc}), we also find that when the $\beta$ in Eq.(\ref{concat_embedding}) is equal to $1$, the prediction effect of SIM-CHGAT is the best, but if the $\beta$ is larger, the prediction performance will decrease to some extent, which can be explained that when the $\beta$ is too large, the behavior of other users will affect the user's own representation of interest. Hence, before the online experiment, we set $\beta$ to $1$.

\end{itemize}





\subsection{Online A/B Experiment}
\label{Online_AB_Experiment}

We apply CHGAT and SIM-CHGAT respectively in the actual online search scenarios of the Koubei app\footnote{https://www.koubei.com}. Consistent with the construction method of offline datasets, we employ the real historical behavior of users to construct heterogeneous graphs and update them in real-time, which is detailed in section\ref{SUPPLEMENT}. Under the framework of the A/B tests, we set one of the buckets as the experimental group and the other bucket as the baseline group and conduct two sets of A/B tests. A/B testS would hit all users who use the search function of the Koubei app. In order to increase the confidence level of the A/B experiment as much as possible, we randomly divide users into buckets with the granularity of days. After 14 days of cumulative testing, the effective data volume in each test bucket is close to two million.

Compared with the online baseline method without CHGAT, the CHGAT proposed in this article has increased the click rate of unique visitors per day(uvCTR) by $3.85\%$, the purchase rate of unique visitors per day(uvCVR) has increased by $2.95\%$, the average revenue per user(ARPU) has been promoted by $7.99\%$, and the average click position of the list of the search result has risen by $4.24\%$. And compared with CHGAT, SIM-CHGAT improves $1.74\%$ on uvCTR, among which the uvCTR of new customers increased by $3.01\%$. These results all show that the algorithm proposed in this article can increase customer flow and overall revenue for merchants on the Koubei platform, and can also help users find more interesting items. In conclusion, our proposed CHGAT has both high application potential and economic value.

\section{Conclusion}

In this paper, in order to solve the challenge that fusing multiple heterogeneous contextual information for dynamic user representation, we propose a context-aware heterogeneous graph attention model to predict user behavior based on dynamic modeling of user interests. Specifically, a variety of different sources of behavioral data is employed to construct heterogeneous graphs of users, and similar crowds behavior graphs are build to solve the problem of limited self-graph scale. Later the proposed unified knowledge representation in CHGAT is able to map multiple vertices in a heterogeneous graph to a similar semantic space. Moreover, the newly designed vertex-level and path-level attention mechanisms are capable of selecting vertices most relevant to outside features in the graph for aggregation. In order to verify the performance of our proposed CHGAT in different scenarios, we employ extensive experiments in large scale offline evaluation datasets and also conduct several two-week online A/B tests. Experimental results demonstrate that the proposed CHGAT achieves obvious advantages compared to other user representation approaches or heterogeneous graph methods and significantly improves the revenue of merchants and the enthusiasm of users on Koubei app. In the future, we will explore more flexible unified knowledge conversion methods.

\bibliographystyle{ACM-Reference-Format}
\bibliography{paper_write}

\appendix

\clearpage

\section{SUPPLEMENT}
\label{SUPPLEMENT}

\begin{figure*}[t] 
\includegraphics[width=0.8\textwidth]{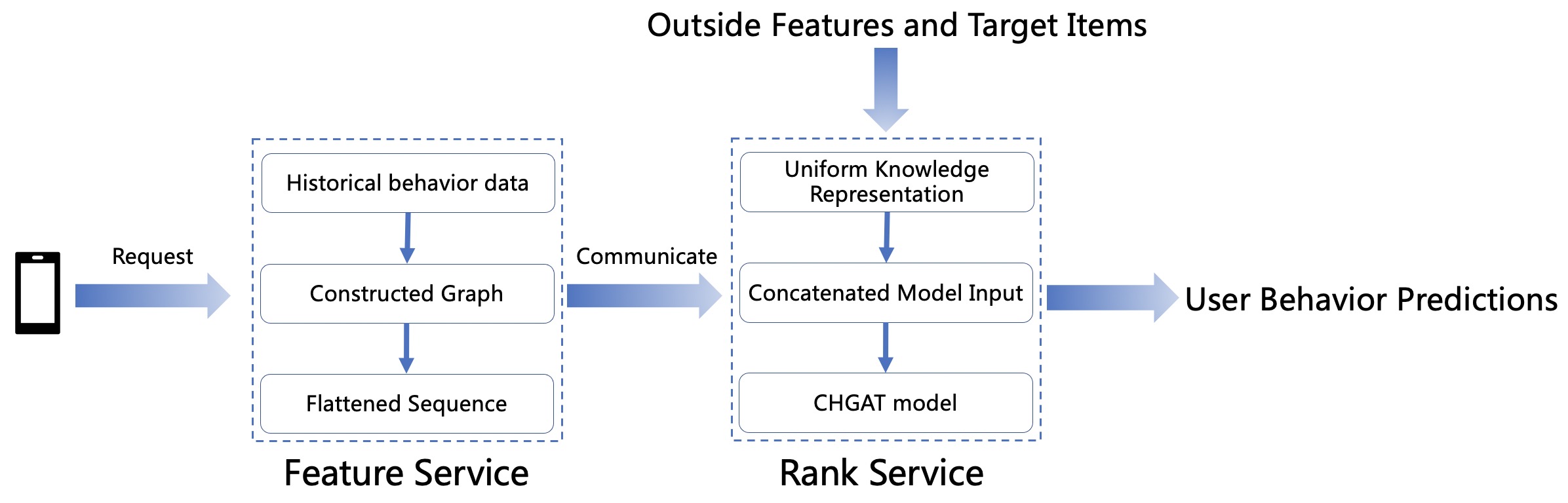} %
\caption{The online service architecture of CHGAT} 
\label{online_structure} 
\end{figure*}

\begin{figure}[b] 
\includegraphics[width=0.25\textwidth]{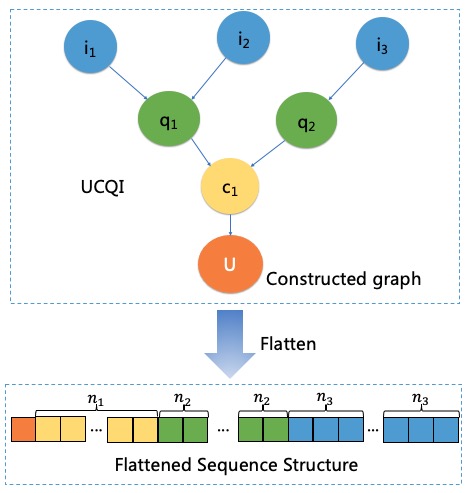} %
\setlength{\abovecaptionskip}{0cm}\caption{The introduction of the flattened sequence} 
\label{sequence} 
\vspace{-2em}
\end{figure}


In order to increase the practicality and reproducibility of the model proposed in this article, we respectively give the data structure and service architecture of online CHGAT in the following subsections.

\subsection{Online Data Structure}

In the modern e-commerce system, the overall time-consuming limitation of the prediction model is usually less than 100ms, or it would affect the feeling of using the e-commerce system. As shown in Fig.(\ref{model_framework}), the graph aggregation process proposed by CHGAT is in a CTR prediction model where real-time requests occur. Directly adopting graph data structure aggregation will cause the model to be unusable online due to high time-consuming.

Therefore, we propose to flatten the graph-like data structure into a sequence structure, and use a fixed position in the sequence to represent the relationship between the nodes in the graph. Take the meta-path $\phi_{UCQI}$ as an example, $n_{1}$ context nodes with the closest occurrence time are reserved, each context node keeps $n_{2}$ corresponding queries, and each query keeps $n_{3}$ interactive items. When the number of nodes in the flattened sequence is insufficient, the default value will be used for completion. As shown in Flg.(\ref{sequence}), let $\mathcal{S}_{\phi_{UCQI}}$ represent the flattened sequence, thus $\mathcal{S}_{\phi_{UCQI}}^{0}$ is the user id and $\mathcal{S}_{\phi_{UCQI}}^{1}$ is the nearest context feature in user history behaviors and $[\mathcal{S}_{\phi_{UCQI}}^{n_{1}+1},\mathcal{S}_{\phi_{UCQI}}^{n_{1}+n_{2}+1}]$ are corresponding queries for context feature in $\mathcal{S}_{\phi_{UCQI}}^{1}$. Similarly, $[\mathcal{S}_{\phi_{UCQI}}^{n_{1}n_{2}+1},\mathcal{S}_{\phi_{UCQI}}^{n_{1}n_{2}+n{3}+1}]$ are corresponding items for query in $\mathcal{S}_{\phi_{UCQI}}^{n_{1}+1}$. When employing online, we only need to obtain data according to the corresponding location to achieve the purpose of high-speed reading of heterogeneous behavior graphs.

\subsection{Online Service Architecture}

In practical applications, we first complete the training of the CHGAT model in the offline environment and deploy it in the online engine to estimate the user's click probability for items to be predicted. The architecture of the online service is shown in the Fig.(\ref{online_structure}). After receiving the user's request, the feature service utilizes the historical behavior database to assemble the user's heterogeneous behavior graph in real-time and flatten it to be a sequence structure. Then in the rank service, the items to be predicted and other context features and the flattened sequence are sent to the unified knowledge representation part. Then the concatenated input tensor can be formulated, and the predicted probability value of the user behavior can be obtained after sending it to the model.

In conclusion, employing the above data structure and service architecture allows our proposed CHGAT to provide accurate prediction results while maintaining the total online service delay below $50ms$, making CHGAT highly practical.



\end{document}